\documentclass[twocolumn]{aastex631}
\usepackage{newtxtext,newtxmath}
\usepackage[T1]{fontenc}
\usepackage{ae,aecompl}
\usepackage{graphicx}
\usepackage{dcolumn}
\usepackage{bm}
\usepackage{color}
\usepackage{natbib}
\usepackage{multirow}

\usepackage{xspace}
\usepackage{nicefrac}
\newcommand{\KEPLER}{{\textsc{Kepler}}\xspace}

\newcommand{\B}[2]{\ensuremath{[\text{#1}/\text{#2}]}\xspace}
\newcommand{\Msun}{\ensuremath{\mathrm{M}_\odot}}

\newcommand{\ch}{\ensuremath{\mathrm{M_{Ch}}}}

\newcommand{\cm}{\ensuremath{\mathrm{cm}}}

\newcommand{\erg}{\ensuremath{\mathrm{erg}}}
\newcommand{\g}{\ensuremath{\mathrm{g}}}

\begin{document}

\title{SDSSJ0018-0939: A Clear Signature of Sub-Chandrasekhar Mass Type 1a Supernova}

\correspondingauthor{S. K. Jeena}
\email{jeenaunni44@gmail.com}

\correspondingauthor{Projjwal Banerjee}
\email{projjwal.banerjee@gmail.com}

\author{S. K. Jeena}
\affiliation{Department of Physics\\
Indian Institute of Technology Palakkad, Kerala, India}

\author{Projjwal Banerjee}
\affiliation{Department of Physics\\
Indian Institute of Technology Palakkad, Kerala, India}

\begin{abstract}
Very metal-poor (VMP) stars (${\rm [Fe/H]}\leq -2$) that have sub-solar values of ${\rm [X/Fe]}$ for $\alpha$ elements such as Mg, Si, and Ca, are referred to as $\alpha$-poor VMP stars. They are quite rare among VMP stars and are thought to have formed from gas enriched predominantly by a single Type Ia supernovae (SN1a) in contrast to most VMP stars which are $\alpha$-enhanced and usually associated with core-collapse supernovae. The observed abundance pattern in such stars can provide a direct way to probe the nucleosynthesis in individual SN1a. Although the abundance patterns in some $\alpha$-poor VMP stars have been shown to be consistent with SN1a ejecta, a clear nucleosynthetic signature for SN1a resulting from the explosion of a near Chandrasekhar mass (near-${\rm M_{Ch}}$) or a sub-Chandrasekhar mass (sub-${\rm M_{Ch}}$) white dwarf, has not been unambiguously detected. We perform a detailed analysis of various formation channels of VMP stars and find that the $\alpha$-poor VMP star SDSSJ0018-0939, which was earlier reported as a star with potential pair-instability supernova origin, provides almost a smoking-gun signature of a sub-${\rm M_{Ch}}$~SN1a resulting from He detonation. We find that compared to other $\alpha$-poor VMP stars that were previously identified with SN1a, SDSSJ0018-0939 is the only star that has a clear and unambiguous signature of SN1a. Interestingly, our results are consistent with constraints on SN1a from recent galactic chemical evolution studies that indicate that sub-${\rm M_{Ch}}$~SN1a account for $\sim 50\hbox{--}75\,\%$ of all SN1a and are possibly the dominant channel in the early Galaxy.

\end{abstract}

\keywords{Chemically peculiar stars(226) --- Stellar nucleosynthesis(1616) --- Type Ia supernovae(1728) --- Population II stars(1284) --- Population III stars(1285)}

\section{Introduction}
Studying very metal-poor (VMP) stars with metallicity $\B{Fe}{H}<-2$ is crucial for understanding the evolution of elements in the early Galaxy during the first billion years after the Big Bang \citep{beers2005,frebel2015}. The majority of VMP stars exhibit an enhanced abundance of $\alpha$ elements such as C, O, Mg, Si, and Ca with super-solar values of $\B{X}{Fe}$ \citep{cayrel2004}. These stars are thought to have formed from gas polluted by core-collapse supernovae (CCSNe), and their detailed abundance pattern can be explained by the mixing and fallback of CCSN ejecta~\citep{umeda2003first, umeda2005variations, heger2010nucleosynthesis, tominaga2014abundance}. On the other hand, VMP stars that have $\B{X}{Fe}<0$ for Mg and most $\alpha$ elements, are known as $\alpha$-poor VMP stars and are quite rare among VMP stars. Their peculiar abundance pattern is typically associated with gas polluted by SN1a~\citep{ivans2003ApJ, Li2022ApJ} as nucleosynthesis in SN1a models naturally produces negligible Mg along with sub-solar values of $\B{X}{Fe}$ for $\alpha$ elements. 

In this regard, it is important to note that the nature of progenitors as well as the mechanism of explosion for SN1a are subjects of ongoing debate, as successful explosions from first principles are still lacking in numerical simulations. Broadly, the scenarios proposed are associated with the explosion of a white dwarf (WD) in a binary configuration with a red giant star (single degenerate) or with another WD (double degenerate). The proposed explosion mechanism includes pure detonation, pure deflagration, and delayed detonation (deflagration to detonation) for near-Chandrasekhar mass (near-\ch) models along with double detonation for sub-Chandrasekhar mass (sub-\ch) models. 
The explosion mechanisms and nucleosynthesis resulting from the various scenarios have been extensively studied over the last several decades ~\citep{nomoto1984, woosley1986, khokhlov1991,iwamoto1999, woosley2011, 2013seitenzahl, fink2014MNRAS, leung2018ApJ, shen2018,  bravo2019A&A, bravo2019MNRAS, leung2020ApJ, lach2020A&A,  2021gronow}. Among the various explosion mechanisms and scenarios that have been proposed, delayed detonation for near-\ch~and double detonation for sub-\ch~SN1a is currently considered to be the dominant channels \citep{shen2018,kirby2019ApJ}. 
The relative frequency of near-\ch~and sub-\ch~models and their contribution to the galactic chemical evolution is one of the crucial questions that several recent studies have tried to constrain. 
By considering the chemical evolution of key elements such as Mn and Ni, several studies have found that a significant fraction of the contribution comes from sub-\ch~SN1a resulting from He detonations accounting for $50\hbox{--}70\,\%$ of all SN1a \citep{seitenzal2013A&A, kirby2019ApJ, reyes2020ApJ, kobayashi2020ApJ, eitner2020, eitner2023}.  Furthermore, some of these also suggest that sub-\ch~SN1a could even be more dominant in the early Galactic evolution with near-\ch~SN1a dominating in the later phases \citep{kirby2019ApJ}. 

An important implication of such findings is that many of the $\alpha$-poor VMP stars should carry a clear signature of individual sub-\ch~SN1a. Detailed abundance in $\alpha$-poor VMP stars can thus be used to confirm this and also provide direct evidence of the existence of sub-\ch~SN1a.  Even though abundance patterns in some $\alpha$-poor VMP stars have been analysed in detail and compared to theoretical SN1a yields, a clear unambiguous match to the observed pattern is still lacking. The situation is made complicated by the fact that a recent study by \citet{jeena_CCSN2024} has pointed out that CCSNe, that do not undergo substantial fallback, can also result in patterns with sub-solar values of $\B{Mg}{Fe}$ that can potentially match the pattern observed in some of the $\alpha$-poor VMP stars. This makes the unambiguous association of an $\alpha$-poor VMP star to an SN1a even more challenging.

We identify a star named SDSSJ0018-0939, which exhibits the most prominent signature of being born from gas polluted by sub-\ch~SN1a resulting from He detonation. SDSSJ0018-0939 is a well-known chemically peculiar VMP star with metallicity $\B{Fe}{H}=-2.5$, which has an elemental abundance pattern that is distinct from other VMP stars and was first reported by~\citet{aoki2014Sci}. 
The peculiar abundance pattern was interpreted by \citet{aoki2014Sci} as a possible signature of a star formed from a gas polluted by ejecta from a pair-instability SN (PISN). However, a detailed analysis of PISN yields by~\citet{takahashi2018ApJ} found that PISN can be ruled out as a source of elements in SDSSJ0018-0939. Our analysis also confirms this where we find that PISN provides a very poor fit to the observed abundance pattern and can be ruled out as the source of elements in SDSSJ0018-0939. Interestingly, however, we find that SDSSJ0018-0939 stands out as a unique star that shows a near smoking-gun signature of a sub-\ch~SN1a.

Our conclusion is based on a novel and detailed analysis that explores the various scenarios of the origin of elements in VMP stars. This includes enrichment from a single PISN, a single CCSN, a combination of ejecta from a single CCSN and a single PISN, a combination of ejecta from two CCSNe, as well as combinations of ejecta from a CCSN and ejecta from both near-\ch~ and sub-\ch~SN1a. We consider a wide range of CCSN models that undergo mixing and fallback along with SN1a from delayed detonation near-\ch~ models and double detonation sub-\ch~models and use a $\chi^2$ minimization to find the best-fit models. 
We find that other $\alpha$-poor VMP stars such as COS171, BD+80245, HE0533-5340, and SMSSJ034249-284216, which have recently been identified with sub-\ch~SN1a \citep{mcwilliam2018ApJ,reggiani2023AJ}, can also be fit well with purely CCSN models and thus cannot be unambiguously associated with SN1a. In sharp contrast, SDSSJ0018-0939 has the clearest signature of a star that is formed from gas polluted dominantly by a sub-\ch~SN1a resulting from a He detonation. We identify key abundance features that make SDSSJ0018-0939 stand out among other $\alpha$-poor VMP stars.

The structure of this paper is as follows: in Sec.~\ref{sec:theor_models}, we discuss the theoretical models used in this study; Sec.~\ref{sec:matching} we discuss the details of the different scenarios for explaining abundance pattern in VMP stars and the methods used to find best-fit models. The results of the analysis for SDSSJ0018-0939 are discussed in Sec.~\ref{sec:best-fit} and the corresponding results for four other $\alpha$-poor VMP stars in Sec.~\ref{sec:other_stars}. Lastly, we finish with the summary and conclusion in Sec.~\ref{sec:summary}.

\section{Theoretical Models and Abundance Patterns} \label{sec:theor_models}
Below, we describe the various theoretical models adopted in this study and highlight the key features of the corresponding elemental abundance patterns.

\subsection{PISN models}\label{sec:PISN_models}
\begin{figure}
    \centering
    \includegraphics[width=\columnwidth]{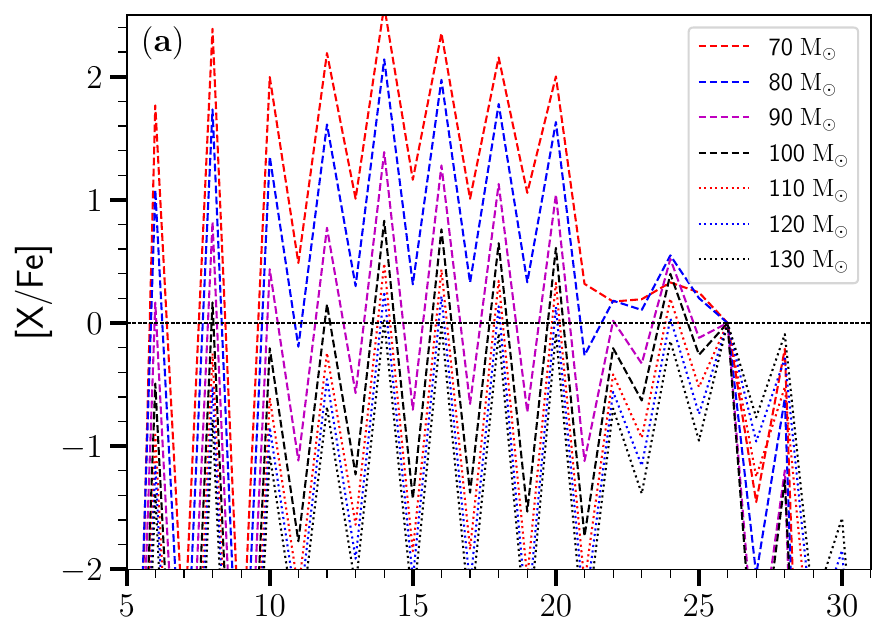} \\
     \includegraphics[width=\columnwidth]{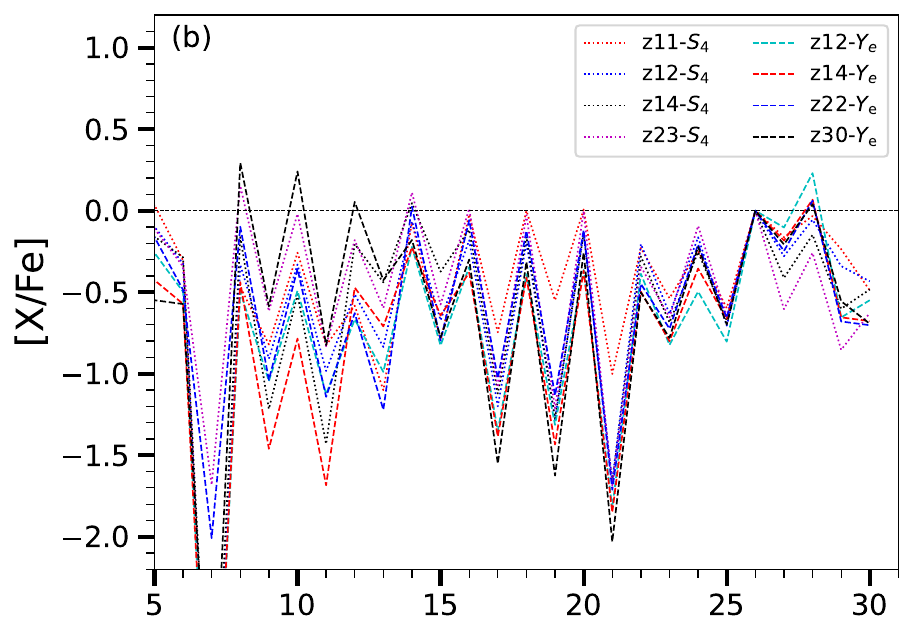} \\
      \includegraphics[width=\columnwidth]{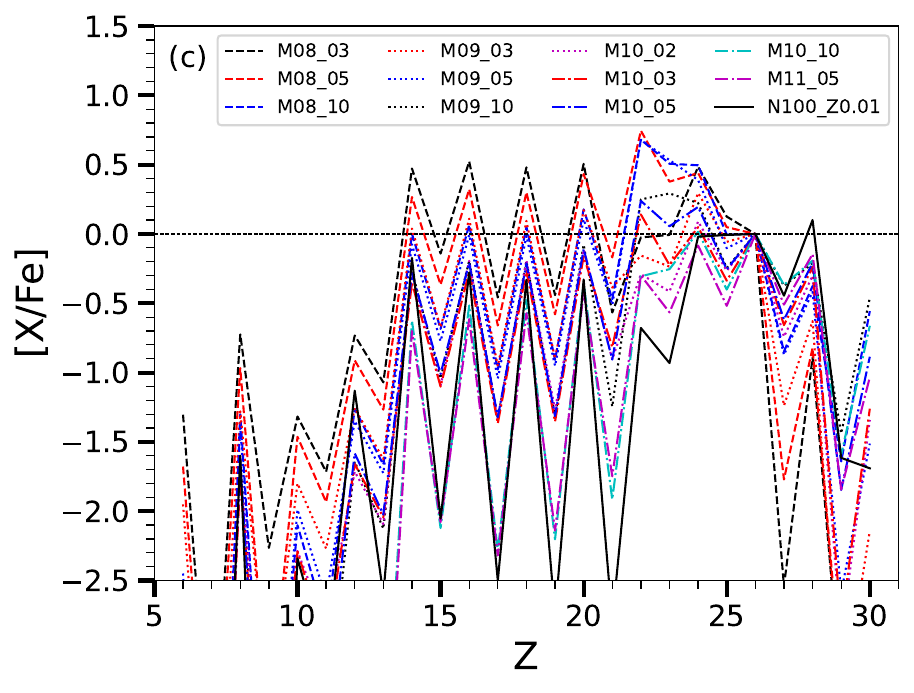}
    \caption{Theoretical elemental pattern relative to Fe from various sources. (a) PISN models of varying He core masses. (b) CCSN ejecta for \texttt{z} models of $10.1$--$30\,\Msun$ without mixing and fallback, and have values of $\B{Mg}{Fe}\lesssim0.0$. (c) SN1a models, where the black solid line corresponds to a typical near-\ch~model, and the dotted and dash-dotted lines represent sub-\ch~models with varied core and shell masses.}
    \label{fig:abundance_pattern}
\end{figure}

We adopt the PISN models from \citet{heger2002} that include the fourteen PISN yields from Pop III stars with He core masses ranging from $65\hbox{--}130\,\Msun$ corresponding to zero-age main sequence mass of $\sim 140\hbox{--}260\,\Msun$. Figure~\ref{fig:abundance_pattern}a shows the corresponding abundance patterns for some of the PISN models covering the progenitor mass range. As can be seen from the figure, the abundance pattern varies substantially with the He core mass.  Lower He core PISN models of $\lesssim 90\,\Msun$ have a large overproduction of light and intermediate elements relative to Fe with a less pronounced odd-even effect. With increasing progenitor mass, the amount of Fe peak production increases dramatically due to explosive Si burning that leads to lower $\B{X}{Fe}$ for elements lighter than Fe along with increased odd-even effect with highly sub-solar values for odd elements.

\subsection{CCSN Models} \label{sec:CCSN_models}
We use a large adaptive co-processing network with reaction rates based on ~\citet{rauscher2002nucleosynthesis} and follow the evolution of the star from birth to death via CCSN using the 1D hydrodynamic stellar evolution code \KEPLER~\citep{weaver1978presupernova, rauscher2003hydrostatic}. We consider non-rotating models of mass ranging from $10.1$--$30\,\Msun$, as detailed in~\citet{jeena_CCSN2024}. The initial composition of these models is taken from~\citet{cyburt2002} that corresponds to primordial Big Bang nucleosynthesis. We refer to these as the \texttt{z} models. 

The explosion is simulated by driving a piston from a mass coordinate, $M_{\rm cut, ini}$, below which all the matter is assumed to fall back to the proto-neutron star. We consider two different choices of $M_{\rm cut, ini}$ in this study, which is identical to \citet{heger2010nucleosynthesis}. In the first case, $M_{\rm cut, ini}$ is chosen to be the mass coordinate where the entropy per baryon $S$ exceeds $4\,k_{\rm B}$, which typically coincides with the base of the oxygen shell. 
In the second case,  $M_{\rm cut, ini}$ is chosen to be the mass coordinate at the edge of the iron core where there is a $Y_{\rm e}$ jump. 
We refer to the former as $S_4$ models and the latter as $Y_{\rm e}$ models. We use a nomenclature where we use the mass followed by the nature of the model ($S_4$ or $Y_{\rm e}$) to clearly identify the various CCSN models. For example, a $15\,\Msun$ progenitor from $S_4$ model is referred to as \texttt{z15}-$S_4$.
We consider explosion energies of  $1.2\times10^{51}\,\erg$ and $1.2\times10^{52}\,\erg$ for all mass models. Lower mass progenitors of $\sim 10$--$12\,\Msun$ are found to have explosion energies of $\lesssim 10^{51}\,\erg$ from detailed 3D simulations~\citep{muller2019MNRAS}. To explore this, we also consider explosion energies of $0.3\times10^{51}\,\erg$ and $0.6\times10^{51}\,\erg$ for all progenitors of initial mass $< 12\,\Msun$.
The ejecta for each SN explosion is calculated using the mixing and fallback model identical to~\citet{jeenaCEMP2023} and ~\cite{jeena_CCSN2024} which are based on the models described in \citet{tominaga2007,ishigaki2014}.  In this model, following the explosive nucleosynthesis by the SN shock, all material above a mass coordinate $M_{\rm cut, fin}$ is fully ejected, whereas a fraction $f_{\rm cut,}$ of the material between $M_{\rm cut, ini}$ and $M_{\rm cut, fin}$ is ejected. $M_{\rm cut, fin}$ and $f_{\rm cut}$, are treated as free parameters, whereas the value of $M_{\rm cut, ini}$ is fixed as mentioned above. 
The value of $M_{\rm cut, fin}$ is varied in steps of $0.1\,\Msun$, starting from a minimum value of $M_{\rm cut, ini}$ to a maximum value that corresponds to the base of the H envelope. The value of $f_{\rm cut}$ varied between 0 to 1. 
Thus, the amount of any isotope ejected by the SN for each model depends on the values of $M_{\rm cut,fin}$ and $f_{\rm cut}$. In this case, $\Delta M_{\rm fb}=(1-f_{\rm cut})\Delta M_{\rm cut}$ is the amount of matter that falls back to the compact remnant, where $\Delta M_{\rm cut}=M_{\rm cut,fin}-M_{\rm cut,ini}$.

The nucleosynthesis of elements in massive stars involves a complex sequence of 
nuclear reactions during various stages of stellar evolution along with explosive nucleosynthesis during SN explosion as discussed in detail in \citet{woosley+2002,heger2010nucleosynthesis} and more recently in ~\citet{jeena_CCSN2024}. The final abundance pattern depends on the explosion energy and mass of the progenitor along with the details of mixing and fallback. However, as shown in ~\citet{jeena_CCSN2024}, sub-solar values of $\B{X}{Fe}$ for $\alpha$ elements result from models with negligible fallback that is particularly relevant for $\alpha$-poor VMP stars. We thus show a few representative CCSN models without fallback that have $\B{Mg}{Fe}\lesssim 0.0$ in Fig.~\ref{fig:abundance_pattern}b. 

\subsection{SN1a Models}\label{sec:SN_1a_models}
In this study, we consider the delayed detonation model for near-\ch~and the double detonation model for sub-\ch~ for SN1a as these are currently considered to be the dominant channels as mentioned in the introduction.
In both cases, we adopt the nucleosynthetic yields obtained from high-resolution 3D studies for the explosion of WDs of low metallicity. In near-\ch~models, we adopt the nucleosynthetic yields derived from the delayed detonation model N100\_Z0.01 by~\citet{2013seitenzahl}, corresponding to a WD with a central density of $2.9\times10^{9}\,\g\,\cm^{-3}$ and an initial metallicity of $0.01\,{\rm Z_\odot}$. For sub-\ch~models, we adopt the yields from eleven double detonation models by ~\cite{2021gronow} that have an initial metallicity of $0.001\,{\rm Z_\odot}$ with CO core masses ranging from $0.8$--$1.1\,\Msun$ that have central densities of $\lesssim 2.5 \times 10^9\,\g\,\cm^{-3}$ with He shell masses of $0.02$--$0.1\,\Msun$. The sub-\ch~models are labelled with the CO core mass ($M_{\rm CO}$) and He shell mass ($M_{\rm He}$), e.g., M09\_05 corresponds to a WD with $M_{\rm CO}=0.9\,\Msun$ and $M_{\rm He}=0.05\,\Msun$.

Figure~\ref{fig:abundance_pattern}c shows the abundance pattern of the adopted SN1a models. In both near-\ch~and sub-\ch~models, there is negligible production of elements lighter than Si. However, the abundance pattern from Si to Fe peak varies significantly among the various sub-\ch~models that are also distinct from the near-\ch~model. In the near-\ch~model, the abundance of $\alpha$ elements is sub-solar ($[\alpha/{\rm Fe}] < 0$) for elements up to Ti (see Fig.~\ref{fig:abundance_pattern}c). 
The sub-\ch~yield patterns, on the other hand, depend on both  $M_{\rm CO}$ and $M_{\rm He}$. The lowest mass $M_{\rm CO}\sim 0.8\,\Msun$ can even produce  $\alpha$-enhanced ($[\alpha/{\rm Fe}>0$) abundance pattern from Si--Cr resulting from incomplete Si burning of both the CO core and He shell. With higher $M_{\rm CO}\gtrsim 1.0\,\Msun$ and $M_{\rm He}\sim 0.1\,\Msun$, the $\alpha$-poor feature, that is usually associated with SN1a, is seen for elements from Si to Ca although Ti to Cr continues to be enhanced with super-solar values of $\B{X}{Fe}$ in some of the models. The enhanced Ti--Cr in such models is exclusively due to the contribution from He shell burning that invariably undergoes incomplete Si burning. 
The contribution of He shell burning to the final ejecta is particularly high for $M_{\rm CO}\lesssim 1\,\Msun$ with intermediate values of $M_{\rm He}\sim 0.05\,\Msun$. Thus, the combination of sub-solar values of $\B{X}{Fe}$ from Si to Ca along with super-solar values from Ti-Cr is a unique feature found only in some of the sub-\ch models that is otherwise not found in either CCSN or near-\ch models. 
Additionally, solar $\B{Mn}{Fe}\sim 0$, usually associated with near-\ch~models, is also seen in the lower CO core mass models whereas, for CO core masses of $\gtrsim 1\,\Msun$, $\B{Mn}{Fe}$ is always sub-solar. Lastly, $\B{Ni}{Fe}$ is sub-solar in all sub-\ch~models which is distinct from near-\ch~model that has $\B{Ni}{Fe}\sim 0.1$.  

In comparison to the CCSN models, the major difference is that SN1a models have negligible production of elements lighter than Si and mostly have sub-solar values of $\B{X}{Fe}$ for $\alpha$ elements from Si to Ca. We note that CCSN models that do not undergo fallback can also have similar features as discussed in Sec.~\ref{sec:CCSN_models}. However, the unique feature seen in some of the sub-\ch~SN1a models from Ti to Cr mentioned above is not found in any of the CCSN models. Even after accounting for mixing and fallback, all CCSN models have sub-solar values of $\B{X}{Fe}$ for Ti to Cr. This is because elements from Ti to Fe are mostly produced in the same region within the CCSN ejecta and are thus insensitive to the details of mixing and fallback. Similarly, $\B{Mn}{Fe}\sim 0$ is not found in any of the CCSN models but is a key signature of the near-\ch~model. We note that $\B{Mn}{Fe}\gtrsim 0$ can also be found in some of the lighter CO core mass sub-\ch~models but the combination of sub-solar values of $\B{X}{Fe}$ for Mg to Ca along with $\B{Mn}{Fe}\sim 0$ is only found in the near-\ch~model.

\section{Matching the Abundance pattern of SDSSJ0018-0939 }\label{sec:matching}
In order to match the observed abundance pattern of SDSSJ0018-0939, we consider six distinct scenarios for the origin of elements in VMP stars formed in the early Galaxy. In the first case, we assume all elements are exclusively produced by a single PISN.
In the second case, we assume all elements are exclusively produced by a single CCSN undergoing mixing and fallback. Next, we consider the mixing of the ejecta from a single CCSN with that of either a near-\ch~or a sub-\ch~SN1a. We then consider the mixing of ejecta of two CCSNe. Finally, we consider the mixing of the ejecta from a single PISN and a single CCSN.

\subsection{Best-fit models from a single PISN}
In order to find the best-fit model, we follow the same procedure as detailed in \citet{jeenaCEMP2023}. For each element ${\rm X}_i$, we define the number yield $Y_{{\rm X}_i}=\Sigma_j m({\rm X}_i^j)/A_j$, where $m({\rm X}^j_i)$ is the yield of the $j^{\rm th}$ isotope and $A_j$ is the corresponding mass number.
The ratio of the number abundance $N$ of any element $X_i$ relative to a reference element $X_{\rm R}$ can be written as
\begin{equation}
    \frac{ N_{{\rm X}_i}}{ N_{\rm   X_R}}=\frac{ Y_{{\rm X}_i}}{ Y_{\rm   X_R}}.
\end{equation}
Following this, the best-fit model parameters are determined using a $\chi^2$ prescription as discussed in \citet{heger2010nucleosynthesis} and more recently in \citet{jeenaCEMP2023, jeena_CCSN2024}. In this prescription, $\chi^2$ is defined as the square of the deviation between the observed and predicted values summed over all elements. The square of the inverse of the observed uncertainty ($\sigma_i$) is used as the weights for each element where we set $\sigma_i=\max(\sigma_i,0.1)$ to ensure that $\chi^2$ is not excessively sensitive to elements with extremely low values of $\sigma_i$. Because PISNe have large explosion energies of $\gtrsim 10^{52}\,\erg$, we impose a minimum dilution of $M{\rm_{ dil,PISN}^{min}}=10^5\,\Msun$ for all models to be consistent with the level of dilution expected from such high energy explosions \citep{chiaki2018}.

\subsection{Best-fit models from a single CCSN}\label{sec:best_fit_ccsn}
The procedure for finding the best-fit model for a single CCSN is similar to the one for PISN described above. In this case, however, both $Y_{{\rm X}_i}$ and $Y_{{\rm X_R}}$ depend on the two free parameters $M_{\rm cut, fin}$ and $f_{\rm cut}$. For every CCSN progenitor, we consider all possible combinations of the two parameters to generate thousands of models. The process is then repeated for each CCSN progenitor. The best-fit model is again determined by the $\chi^2$ formulation mentioned above. 
In order to be consistent with the dilution found in simulations of metal mixing from CCSN, we only consider solutions for which the effective dilution mass from CCSN is greater than $ M{\rm _{dil,CCSN}^{min} }= 10^4\,\Msun$ for explosion energies of $\leq 1.2\times 10^{51}\,\erg$ and $ M{\rm _{dil,CCSN}^{min} }= 10^5\,\Msun$ for $1.2\times 10^{52}\,\erg$. These values correspond roughly to the minimum value of the effective dilution mass for the CCSN ejecta from the detailed simulation of inhomogeneous metal mixing in the early Galaxy~\citep{chiaki2018, magg2020minimum}. 

We note here that detailed 3D simulations have shown that Sc is dominantly produced by neutrino-processed proton-rich ejecta that are not modelled in our 1D calculations. Sc yield in 3D models, that include detailed neutrino interaction, can be a factor of $\gtrsim 10$ higher than the corresponding 1D model \citep{siverding2020ApJ,wang2024ApJ} at solar metallicity. This implies that almost all of the Sc at zero metallicity will be from the neutrino-processed material. Because of this reason, in our analysis, we treat the measured abundance of Sc (if available) as an upper limit. The same is true for K which has negligible production in the ejecta of Pop III stars but will be produced entirely in the neutrino-processed material that is not included in our models. We also treat K as an upper limit whenever they are detected.

\subsection{Best-fit models from the mixing of CCSN and SN1a ejecta} \label{subsec:mixing_CCSN+SN1a}
In order to determine the best-fit model based on the mixing of ejecta from two distinct sources, we consider the prescription used in the recent work~\citet{jeenaCEMP2023}. In this prescription, the mixing of the ejecta of the two sources, i.e.,  CCSN and SN1a, is parameterized by a single parameter $\alpha$ given by
\begin{equation}
    \alpha =\frac{ M{\rm_{dil,CCSN}}}{M{\rm_{dil,CCSN}}+M{\rm_{dil,1a}}},
    \label{eq:alpha}
\end{equation}
where $M{\rm_{dil,CCSN}}$ and $M{\rm_{dil,1a}}$ are the effective dilution masses from CCSN and SN1a, respectively. 
Following the mixing procedure, the number abundance of any element ${\rm X}_i$ relative to a reference element ${\rm X_R}$ can be written as 
\begin{equation}
    \frac{ N_{{\rm X}_i}}{ N_{\rm   X_R}}=\frac{ Y_{{\rm X}_i}}{ Y_{\rm   X_R}}=\frac{{ \alpha Y_{{{\rm X}_i},{\rm 1a}}+\left(1-\alpha\right)\, Y_{{{\rm X}_i},{\rm CCSN}}(M_{\rm cut,fin},f_{\rm cut})}}{{ \alpha Y_{\rm{X_R,1a}}+\left(1-\alpha\right)\, Y_{\rm X_R,CCSN}(M_{\rm cut,fin},f_{\rm cut})}},
\end{equation}
where the subscript `1a' and `CCSN' are used for the number yields corresponding to SN1a and CCSN, respectively. 
In this case, the final abundance pattern depends on the parameter $\alpha$ in addition to $M_{\rm cut,fin }$ and $f_{\rm cut}$. The three parameters are then varied to find the best-fit model using the same $\chi^2$ prescription used for PISN and CCSN. For any given fit, the relative contribution from each source to an element ${\rm X}_i$, can be quantified by computing the fraction $\eta ({\rm X}_i)$ of the total elemental yield $Y_{{\rm X}_i}$. We compute $\eta ({\rm X}_i)$ for SN1a and CCSN, i.e., $\eta_{\rm 1a}({\rm X}_i)$ and $\eta_{\rm CCSN}({\rm X}_i)$, where  $\eta_{\rm 1a}({\rm X}_i)\,+\,\eta_{\rm CCSN}({\rm X}_i)=1$.

Depending on the type of SN1a, we consider two different mixing scenarios:
\begin{itemize}
    \item  CCSN+near-\ch: we consider mixing of CCSN ejecta with the ejecta from the near-\ch~model from~\citet{2013seitenzahl}. 
    
    \item CCSN+sub-\ch: we consider mixing of CCSN ejecta with the ejecta from sub-\ch~models from~\citet{2021gronow} for the various CO core and He shell masses as discussed in Sec.~\ref{sec:SN_1a_models}.
\end{itemize}
In both cases, we consider mixing and fallback of CCSN models as discussed in Sec.~\ref{sec:CCSN_models}.  
Thus,  CCSN+near-\ch~and CCSN+sub-\ch~include the mixing between all the combinations of CCSN ejecta from mixing and fallback with SN1a ejecta. The mixing parameter $\alpha$ controls the mixing of the CCSN and SN1a ejecta, where the values are varied from a lowest value of $10^{-7}$ (negligible SN1a contribution) to a maximum value of $1$ (exclusively SN1a contribution). In both scenarios, we find the best-fit models by minimising $\chi^2$ as mentioned earlier, where we again impose the criteria that both $M{\rm_{dil,1a}}$ and $M{\rm_{dil,CCSN}}$ are greater than a minimum dilution, where we adopt $M{\rm_{dil,1a}^{min}}=10^4\,\Msun$ and the same values of $ M{\rm _{dil,CCSN}^{min} }$ as mentioned in Sec.\ref{sec:best_fit_ccsn}. Similar to the single CCSN model, we treat the observed values of Sc and K as upper limits. 

\begin{figure*}
    \centering
    \includegraphics[width=\textwidth]{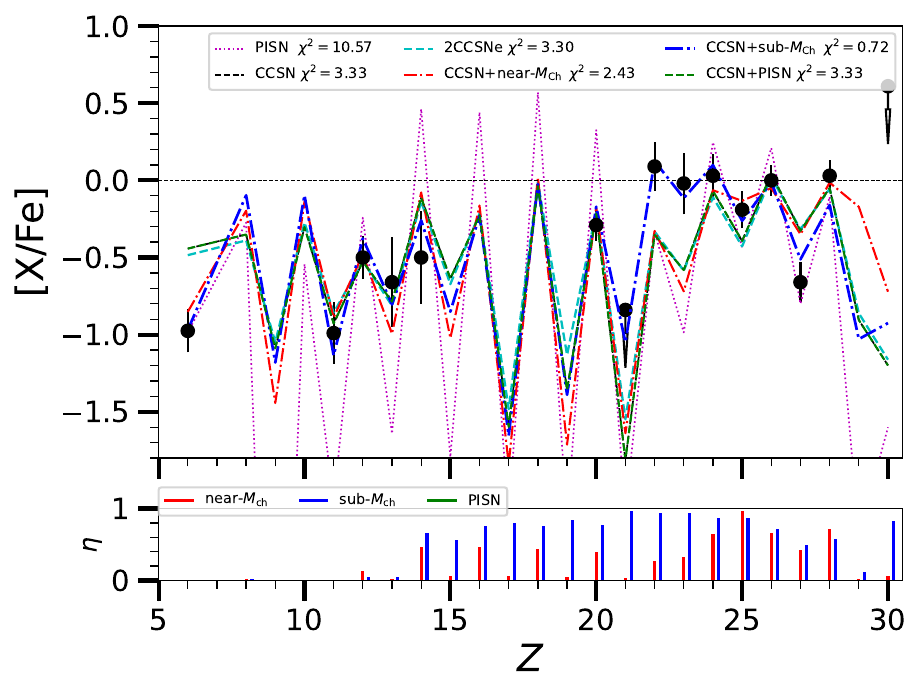}
    \caption{Top: The elemental abundance pattern of SDSSJ0018-0939 compared with the best-fit models from various scenarios: PISN (magenta dotted line), CCSN (black dashed line), 2CCSNe (cyan dashed line),  CCSN+near-\ch~ (red dash-dotted line), CCSN+sub-\ch~(blue dash-dotted line), and CCSN+PISN (green dashed line). Bottom: The fraction $\eta$ for all elements produced by near-\ch~in CCSN+near-\ch~scenario (red), sub-\ch~in CCSN+sub-\ch~scenario (blue), and PISN in CCSN+PISN scenario (green).}
    \label{fig:SD0018_best-fit_comp}
\end{figure*}

\subsection{Best-fit models from the mixing of ejecta from two CCSNe}
It is important to note that because $\alpha\approx0$ essentially corresponds to zero contribution from SN1a, implies that all possible single CCSN models are subsets of the set of models for the CCSN+SN1a scenario. This guarantees that the $\chi^2$ from the best-fit CCSN+SN1a is always lower than the $\chi^2$ from a single CCSN scenario. 
Thus, in order to evaluate whether an observed abundance pattern can be fit better by CCSN relative to CCSN+SN1a, we consider the possibility of mixing the ejecta from two different CCSN. Unfortunately, each CCSN progenitor has thousands of possible ejecta corresponding to the different combinations of $M_{\rm cut,fin}$ and $f_{\rm cut}$.  This makes the total number of ejecta models, from all combinations of parameters for combining the ejecta from two CCSN, exceedingly large ($\gtrsim 10^{11}$ models) and computationally prohibitively expensive. We thus explore the limited parameter space of combining the ejecta from mixing and fallback from a CCSN with that of another CCSN that undergoes no fallback, i.e., $f_{\rm cut}=1$. Although this explores a tiny fraction of the total parameter space of the mixing between two CCSN, the combinations cover the most important part relevant for $\alpha$-poor VMP stars that tend to prefer CCSN models with negligible fallback as fallback invariably increases $[\alpha/{\rm Fe}]$. For brevity, we refer to this scenario as 2CCSNe from hereon.
In order to find the best-fit model, we follow the same procedure as used for the mixing between CCSN and SN1a as discussed in Sec.~\ref{subsec:mixing_CCSN+SN1a} where SN1a is replaced by the second CCSN ejecta with no fallback. As in the case of the single CCSN scenario, we treat the observed values of Sc and K as upper limits. 

\subsection{Best-fit models from the mixing of ejecta from a PISN and a CCSN}
Finally, we consider the mixing of the ejecta between a single PISN and a single CCSN model. The procedure is identical to the one listed in Sec.\ref{subsec:mixing_CCSN+SN1a}. Similar to the CCSN+SN1a scenarios, the CCSN models explore all possible combinations of mixing and fallback for each mass model that is combined with all possible PISN models.
Similar to single PISN and single CCSN scenarios, we impose the same minimum dilution mass for all models and treat Sc and K as upper limits instead of using the observed values. 
\section{Best-fit Models for SDSSJ0018-0939}\label{sec:best-fit}

\begin{table*}
\centering
    \caption{Best-fit models and parameters along with the dilution mass for SDSSJ0018-0939. }
     \resizebox{\linewidth}{!}{%
    \begin{tabular}{cccccccc}
    \hline
    Scenario&Model name&$E_{\rm exp}$ &$\chi^2$&$\alpha$&$\Delta M_{
    \rm cut}$&$\Delta M_{\rm fb}$&$M{\rm_{dil}}$\\
    & & ($\times 10^{51}\,\erg$)& &&(\Msun) &(\Msun)& ($\times 10^4\,\Msun$)\\
    \hline
       PISN& $120\,\Msun$ He core&71.0&10.57&--&--&--&$4.0\times 10^4$\\ 
         CCSN& \texttt{z11.8}-$Y_{\rm e}$ &1.2&3.33&--&0.11&0.07&1.91\\ 
         
         \multirow{2}{*}{CCSN+PISN}&\texttt{z11.8}-$Y_{\rm e} $ +      &1.2& \multirow{2}{*}{3.33}& \multirow{2}{*}{$10^{-7}$}&0.11 &0.07  &1.9\\
                                   &$120 \,\Msun$ He core              &71.0& &                                               &--   &--    &$1.9 \times 10^7$\\
         \multirow{2}{*}{2CCSNe} & \texttt{z11.3}-$Y_{\rm e} +$&1.2&  \multirow{2}{*}{3.30}& \multirow{2}{*}{0.25}&0.00&0.00&7.85 \\
         & \texttt{z11.8}-$Y_{\rm e} $& 1.2& &&0.11& 0.06&2.62\\ 
          \multirow{2}{*}{CCSN+near-\ch}&\texttt{z20}-$S_4 +$&12.0 & \multirow{2}{*}{2.43}& \multirow{2}{*}{0.33}&1.07 & 0.00&13.70\\ 
         &  N100\_Z0.01&-- & && --&--&27.80 \\ 
         \multirow{2}{*}{CCSN+sub-\ch}&\texttt{z30}-$Y_{\rm e} +$& 1.2& \multirow{2}{*}{0.72}& \multirow{2}{*}{0.57}&4.23&1.60&24.60\\
         &   M10\_05& --&& &-- &--&18.50 \\

        \hline
    \end{tabular}}
    
    \label{tab:best-fit_parm}
\end{table*}

Figure~\ref{fig:SD0018_best-fit_comp} shows the best-fit models from all possible combinations, i.e., exclusively from a single PISN (magenta dotted line),  a single CCSN (black dashed line), 2CCSNe (cyan dashed line), CCSN+near-\ch~SN1a (red dash-dotted line), CCSN+sub-\ch~SN1a (blue dash-dotted line), and CCSN+PISN (green dashed line).
The fraction $\eta$ for all elements produced by near-\ch in CCSN+near-\ch scenario (red), sub-\ch in CCSN+sub-\ch scenario (blue), and PISN in CCSN+PISN scenario (green) is also shown in the bottom panel of Fig.~\ref{fig:SD0018_best-fit_comp}. The best-fit models and the corresponding parameters are listed in Table~\ref{tab:best-fit_parm}.

As can be seen from Fig.~\ref{fig:SD0018_best-fit_comp}, the best-fit PISN model provides a very poor fit with $\chi^2=10.57$ and fails to fit multiple elements with particularly large deviation from the observed values for Na, Al, Si, Ca, Ti and V. Compared to PISN models, the best-fit CCSN model provides a better fit with $\chi^2=3.33$, and can fit most of the elements from Na to Ca and Cr to Ni. However, even after accounting for the $1\sigma$ uncertainty, it fails to fit the observed abundances of several elements such as C, Si, Ti, V, Mn, and Co with large deviations of $\sim 0.5\,{\rm dex}$ from the observed values for C, Ti, and V. The best-fit CCSN+PISN model provides the exact same solution as the best-fit single CCSN with effectively zero contribution from PISN with $\eta_{\rm PISN}\lesssim 10^{-4}$ ($\lesssim 0.01\%$) for all elements confirming the lack of any discernible PISN features in the star.   
The situation does not improve when mixing from 2CCSNe is considered with a negligible improvement in the quality of fit with a $\chi^2=3.30$. The particularly poor fit for C, Ti, and V for the best-fit models from both single CCSN and 2CCSNe scenarios strongly indicates that the observed pattern cannot be explained by CCSN ejecta alone.
Here, it is important to note that because of the high value of $\log g=5$, the highly sub-solar value of $\B{C}{Fe}\sim -1$ cannot be attributed to the destruction of C due to internal mixing in SDSSJ0018-0939. Thus, the very low $\B{C}{Fe}$ in this star is inherited from the birth material and a strong indicator of additional Fe contribution from SN1a. This can be seen clearly from the best-fit CCSN+near-\ch~model that provides a slightly better fit with a lower $\chi^2=2.43$ which is largely due to a much better fit to C. Compared to CCSN models, higher [Mn/Fe] from near-\ch~models leads to a perfect fit for Mn, but is counterbalanced by a worse fit to Al. Similar to the best-fit CCSN models, the best-fit CCSN+near-\ch~model cannot fit C, Ti, V, and Co, with a particularly large deviation from the observed values for Ti and V.  

In contrast, the best-fit model from the CCSN+sub-\ch, resulting from the mixing of the ejecta from a $30\,\Msun$ CCSN and a sub-\ch~SN1a with $M_{\rm CO}=1\,\Msun$ and $M_{\rm He}=0.05\,\Msun$, provides an excellent fit to the observed abundance pattern with the lowest $\chi^2=0.72$. 
In particular, the best-fit CCSN+sub-\ch~model can explain almost all the elements except for Ni and can perfectly fit C along with all elements from Ti--Fe.  
Importantly, the near perfect fit for Ti to Cr is entirely due to the contribution for sub-\ch~SN1a as is evident from the fact that $\eta_{\rm 1a}\sim 1$ (see the bottom panel of Fig.~\ref{fig:SD0018_best-fit_comp}) from Sc to Mn for the best-fit model. We note that the particular feature of $\B{X}{Fe}\gtrsim 0$ for Ti, V, and Cr with $\B{Ti}{Cr}\gtrsim 0$ can only be fit by sub-\ch~models that arise from the incomplete Si burning of the He shell as discussed earlier in Sec.~\ref{sec:SN_1a_models}.
Ejecta from sub-\ch~SN1a also contributes majorly to all elements from Si--Ca as well as Fe--Ni with $\eta_{\rm 1a}\gtrsim 0.5$. The CCSN ejecta, on the other hand, accounts for almost all the elements lighter than Si but is a subdominant contributor to elements from S--Fe. For elements such as Si, Co, and Ni, both CCSN and  sub-\ch~ejecta make comparable contributions. Because the contribution from CCSN ejecta to the Fe group is non-negligible, this forces the CCSN model to have a very low fallback of the innermost material containing Fe group elements in order to fit the sub-solar values of $\B{X}{Fe}$ from C to Si. 

It is important to note that in the CCSN+sub-\ch~scenario, although the contribution of both CCSN and sub-\ch~SN1a is required to fit the full observed abundance pattern, the primary reason for the excellent fit is driven by the crucial contribution of sub-\ch~SN1a that can fit the unique pattern from Ti--Cr. We find that in addition to the $30\,\Msun$ CCSN model that provides the best fit, many other CCSN models that have a low fallback give comparable fits with very low $\chi^2\lesssim 0.8$ when their ejecta is mixed with the sub-\ch~ SN1a model. Similar to the best-fit CCSN+sub-\ch~model, all such models have $\eta_{\rm 1a}\gtrsim 1$ for elements from Ti--Cr, confirming the fact that the primary reason for the excellent fit is due to  sub-\ch~ SN1a contribution.

\begin{figure*}
\centering
    \includegraphics[width=0.47\linewidth]{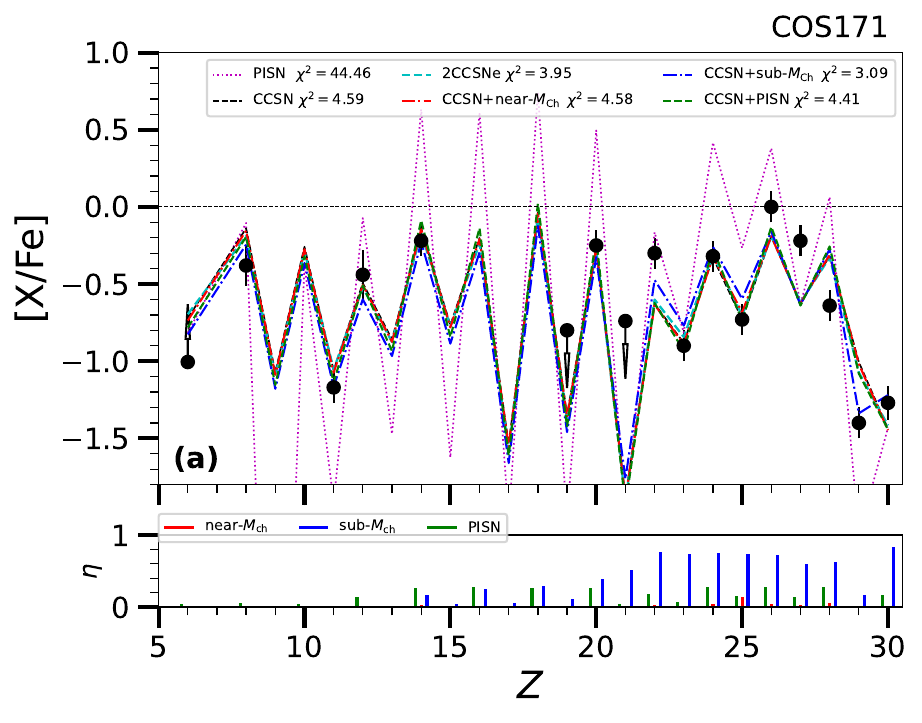}\hfil
    \includegraphics[width=0.47\linewidth]{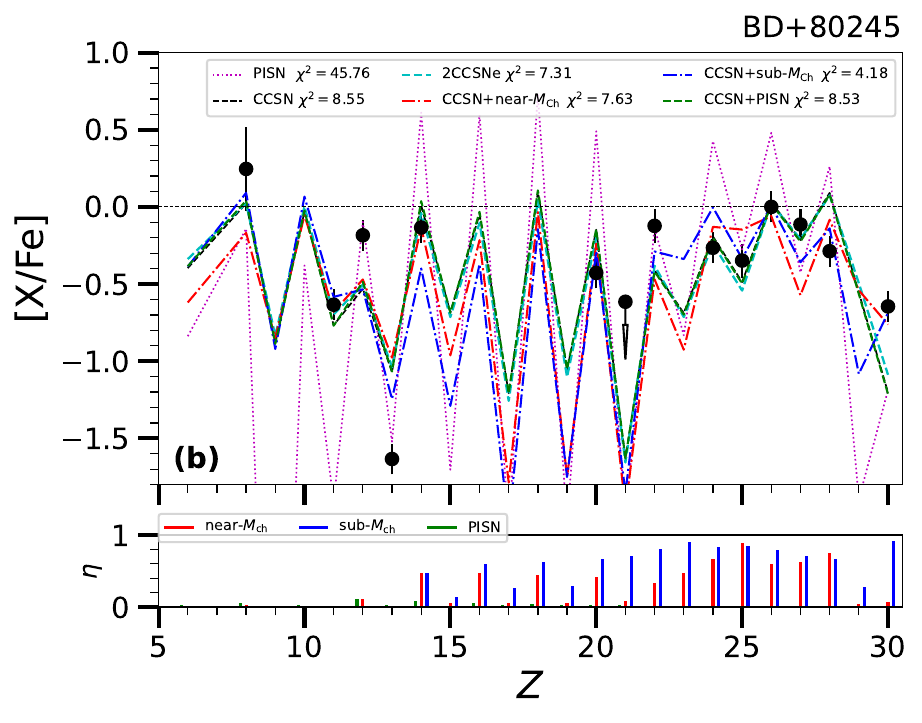}\par\medskip
    \includegraphics[width=0.47\linewidth]{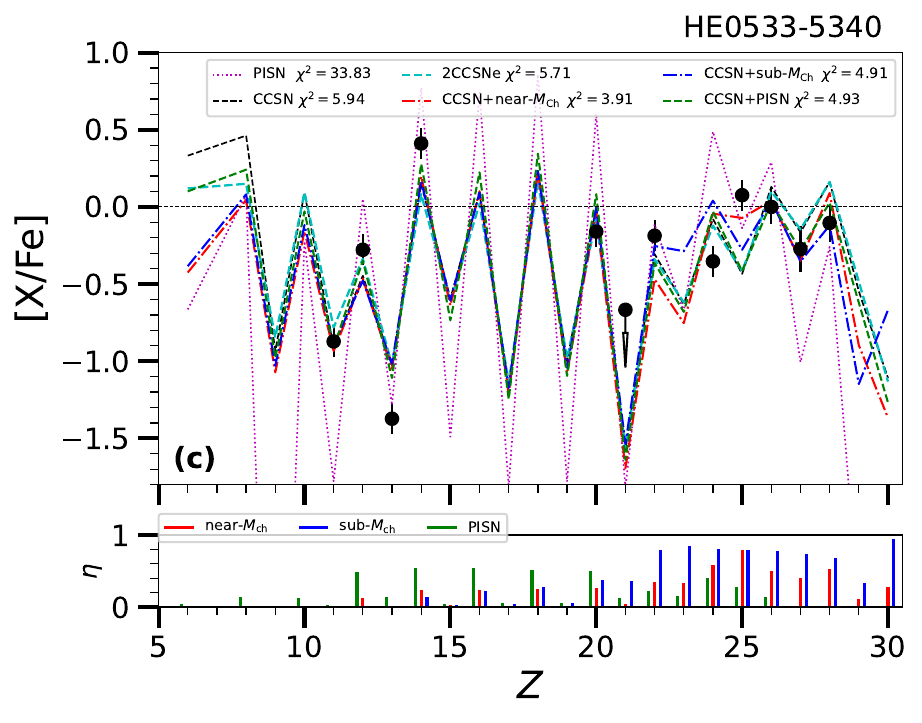}\hfil
    \includegraphics[width=0.47\linewidth]{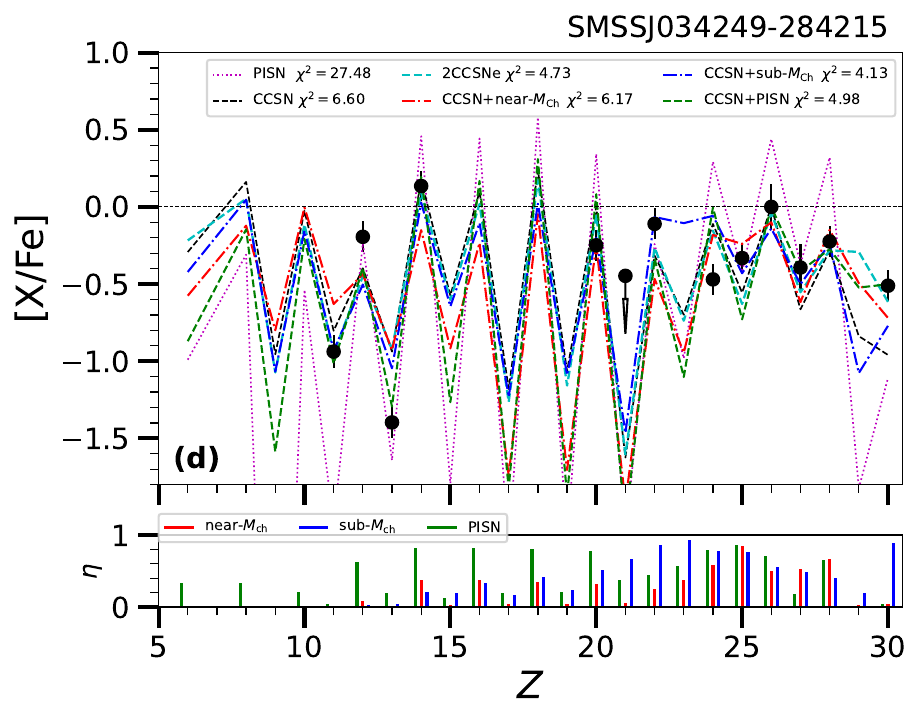}
\caption{Same as Fig.~\ref{fig:SD0018_best-fit_comp}, but for stars COS171, BD+80245, HE0533–5340, and SMSSJ034249–284216. 
}
\label{fig:best-fit_stars}
\end{figure*}

\begin{table*}
    \centering
    \caption{Same as Table.~\ref{tab:best-fit_parm}, but for stars COS171, BD+80245, HE0533-5340, and SMSSJ03429-284216. }
     \resizebox{\linewidth}{!}{%
    \begin{tabular}{ccccccccc}
    \hline
    Star & Scenario&Model name&$E_{\rm exp}$ &$\chi^2$&$\alpha$&$\Delta M_{\rm cut}$&$\Delta M_{\rm fb}$&$M{\rm_{dil}}$\\
    & && ($\times 10^{51}\,\erg$)& & &(\Msun)&(\Msun)& ($\times 10^4\,\Msun$)\\
    \hline
        \multirow{10}{*}{COS171}&PISN & $120\,\Msun$ He core& 71.0& 44.46&-- &--&-- &$2.1\times10^3$\\ 
        &CCSN& \texttt{z23}-$Y_{\rm e}$& 1.2& 4.59&--&0.11& 0.11& 1.10\\
         &\multirow{2}{*}{CCSN+PISN}& \texttt{z23}-$Y_{\rm e} + $  &1.2& \multirow{2}{*}{4.41}& \multirow{2}{*}{0.004}&0.11 &0.11 &$1.33$\\
         &                          &$130 \,\Msun$ He core         &87.3& &                                           &--   &--   & $3.3 \times 10^2$\\
         &\multirow{2}{*}{2CCSNe}& \texttt{z11.6}-$Y_{\rm e} +$ & 1.2&  \multirow{2}{*}{3.95}& \multirow{2}{*}{0.57}&0.00&0.00 & 1.35\\ 
         & & \texttt{z23}-$Y_{\rm e}$&1.2 & &&0.21&0.20&1.02\\ 
       &2CCSNe &\texttt{z11.5}-$Y_{\rm e} +$& 1.2& \multirow{2}{*}{3.36}& \multirow{2}{*}{0.50}&0.00&0.00 &0.50\\ 
       &($M_{\rm dil,CCSN}^{\rm min}=5\times 10^3\,\Msun$) &\texttt{z14.1}-$Y_{\rm e}$& 1.2&&&0.20&0.19&0.50\\ 
        & \multirow{2}{*}{CCSN+near-\ch}&\texttt{z23}-$Y_{\rm e} +$&1.2 & \multirow{2}{*}{4.58}& \multirow{2}{*}{0.02}&0.11&0.11 &1.14\\ 
         &&N100\_Z0.01&-- && &-- &--&55.70\\ 
         &\multirow{2}{*}{CCSN+sub-\ch}&\texttt{z23}-$Y_{\rm e} +$& 1.2& \multirow{2}{*}{3.09}& \multirow{2}{*}{0.32}&0.68&0.47 &1.41\\
         &&M11\_05  &-- & &&--&--&2.99  \\ 
         \hline
         \multirow{11}{*}{BD+80245}& PISN& $125\,\Msun$ He core &78.8 &45.7& --& --&--&$4.3\times10^3$\\ 
         & CCSN & \texttt{z20.5}-$Y_{\rm e}$& 1.2&8.55& --&1.45&0.29 & 1.00\\ 
         &($M_{\rm dil,CCSN}^{\rm min}=5\times10^3\,\Msun$) &\texttt{z20.5}-$Y_{\rm e}$ & 1.2&8.37&--&2.11 &1.28&0.51 \\ 
         &\multirow{2}{*}{CCSN+PISN}& \texttt{z20.5}-$Y_{\rm e} +$      &1.2 & \multirow{2}{*}{8.53}& \multirow{2}{*}{0.002}&1.54   &0.34  &1.00\\
         &                          &$70 \,\Msun$  He core              &8.9& &                                             &--   &--      &$5.0 \times 10^2$\\
         &\multirow{2}{*}{2CCSNe}&\texttt{z10.1}-$Y_{\rm e} +$ &1.2& \multirow{2}{*}{7.31}& \multirow{2}{*}{0.50}&0.00&0.00 & 1.00\\
         & & \texttt{z20.5}-$Y_{\rm e}$&1.2 &&&1.45&0.51&1.00\\  
         
         &2CCSNe &\texttt{z10.1}-$Y_{\rm e} +$&1.2& \multirow{2}{*}{6.58}& \multirow{2}{*}{0.50}&0.00&0.00 &0.51\\
         &($M_{\rm dil,CCSN}^{\rm min}=5\times10^3\,\Msun$)&\texttt{z20.5}-$Y_{\rm e}$& 1.2&&&2.21&1.63&0.51\\ 
         & \multirow{2}{*}{CCSN+near-\ch}&\texttt{z17.2}-$S_{\rm 4} +$ &1.2& \multirow{2}{*}{7.63}& \multirow{2}{*}{0.19}& 1.26&0.03&1.09\\ 
          &&N100\_Z0.01&-- & & &-- &--&4.64\\ 
          &\multirow{2}{*}{CCSN+sub-\ch}& \texttt{z28}-$Y_{\rm e} +$& 1.2& \multirow{2}{*}{4.18}& \multirow{2}{*}{0.24}&4.53& 3.67&1.01\\
          &&  M10\_10& --& &&-- &--&3.18 \\ 
           
          \hline
         \multirow{8}{*}{HE0533-5340}& PISN& $110\,\Msun$ He core& 56.4& 33.83& --&-- &--& $1.72\times10^4$ \\ 
         & CCSN&\texttt{z22}-$Y_{\rm e}$&1.2&5.94&-- & 3.08&2.68& 1.05\\
         &\multirow{2}{*}{CCSN+PISN}&\texttt{z20.5}-$Y_{\rm e} $ +     &1.2   & \multirow{2}{*}{4.93}& \multirow{2}{*}{0.002}&2.59   &2.15  &1.67\\
         &                          &$100 \,\Msun$ He core             &41.9& &                                              &--     &--      & $8.3 \times 10^2$\\
         &\multirow{2}{*}{2CCSNe}&\texttt{z20.5}-$Y_{\rm e} +$&1.2& \multirow{2}{*}{5.71}& \multirow{2}{*}{0.13} & 0.00&0.00 & 1.07\\ 
         &&\texttt{z13.2}-$Y_{\rm e}$&1.2 &&&0.60&0.59& 7.16\\ 
         &  \multirow{2}{*}{CCSN+near-\ch}&\texttt{z22}-$Y_{\rm e}$ &1.2& \multirow{2}{*}{3.91}&  \multirow{2}{*}{0.23}&0.68& 0.40& 8.54\\
         &&N100\_Z0.01& --& &&--&--&28.60 \\ 
         &\multirow{2}{*}{CCSN+sub-\ch}&\texttt{z22}-$Y_{\rm e}$& 1.2& \multirow{2}{*}{4.91}& \multirow{2}{*}{0.28}&0.68&0.58 & 7.77\\
         && M09\_03& --& & &-- &--&20.00\\ 
        \hline
         \multirow{8}{*}{SMSSJ034249-284216}&PISN&$130\,\Msun$ He core&87.3 & 27.48&-- &--&-- & $1.3\times10^4$\\  
         & CCSN&\texttt{z22}-$S_{\rm 4} $&1.2&6.60& --&1.74& 0.00&2.00\\ 
         &\multirow{2}{*}{CCSN+PISN}&\texttt{z27}-$Y_{\rm e} $ +      &12.0 & \multirow{2}{*}{4.98}& \multirow{2}{*}{0.04}&3.10  &0.96  &10.0\\
         &                          &$120 \,\Msun$ He core            &71.0& &                                            &--    & --   & $2.4 \times 10^2$\\
         &\multirow{2}{*}{2CCSNe}&\texttt{z22}-$S_{\rm 4} +$ &1.2& \multirow{2}{*}{ 4.73}& \multirow{2}{*}{ 0.24}& 0.00&0.00& 1.01\\
         & &\texttt{z12.7}-$S_{\rm 4} $&1.2 & &&0.40&0.18&3.19\\ 
         &  \multirow{2}{*}{CCSN+near-\ch}&\texttt{z17.2}-$S_{\rm 4} +$ &1.2& \multirow{2}{*}{6.17}& \multirow{2}{*}{0.14}&0.88& 0.00& 2.10\\
         &&N100\_Z0.01&-- & & &--&--&12.80 \\ 
         &\multirow{2}{*}{CCSN+sub-\ch}&\texttt{z22}-$Y_{\rm e}$&1.2& \multirow{2}{*}{4.13}& \multirow{2}{*}{0.23}&0.88& 0.66 & 2.70 \\
         && M10\_10 &-- && &-- &--&9.10 \\ 
         \hline
         
    \end{tabular}}
    
    \label{tab:best-fit_parm_stars}
\end{table*}

\section{Signature of SN1a in Other Stars?}\label{sec:other_stars}

As mentioned in the introduction, although $\alpha$-poor VMP stars have been associated with SN1a, very few studies have attempted to identify the observed pattern in individual stars with theoretical yields from SN1a. An exception to this has been the work by \citet{mcwilliam2018ApJ} where conclusive evidence of sub-\ch~SN1a origin for the star COS171 in Ursa Minor was reported. Instead of considering the contribution of CCSN directly, the abundance of a reference VMP star in Ursa Minor was used as a proxy for the composition of the gas polluted by CCSN which was then combined with SN1a ejecta.  
Recently, \citet{reggiani2023AJ} has also identified three $\alpha$-poor stars that show possible signatures of sub-\ch~SN1a contribution. However, only pure SN1a ejecta was considered in the analysis and CCSN contribution was neglected. Among the three stars, BD+80245 has the highest metallicity of $\B{Fe}{H}=-1.73$ whereas SMSSJ034249–284216 and HE 0533–5340 have metallicities of $\B{Fe}{H}=-1.97$ and $-2.44$, respectively.  Below we analyse these stars for the different scenarios considered for SDSSJ0018-0939 to see if the observed abundance patterns can be explained by PISN or CCSN, or whether SN1a contribution from either near-\ch~ or sub-\ch~ models is required.

\subsection{COS171}
COS171 is an $\alpha$-poor star with $\B{Fe}{H}=-1.35$ which happens to be one of the highest metallicity stars in Ursa Minor~\citep{cohen2010ApJ}. Because this star has a low $\log g=0.8$, the initial C in the star would have been considerably depleted \citep{Placco2014Apj}.  For this reason, we treat the observed value of C in this star as a lower limit.  We analysed this star for all the possible scenarios and the best-fit models are shown in Fig.~\ref{fig:best-fit_stars}a and the corresponding details are listed in Table~\ref{tab:best-fit_parm_stars}. 

The best-fit PISN model provides an extremely poor fit with $\chi^2=44.46$ and clearly cannot explain the abundance pattern. For a single CCSN, the best-fit model is from a $23\,\Msun$ progenitor with negligible fallback with a $\chi^2=4.59$  and can fit all elements from Na to Ca and  V to Mn. However, the quality of fit is particularly poor for Ti and all elements from Fe to Cu as the fit is well beyond the observed $1\sigma$ uncertainty. Additionally, it fails to fit O and Zn. The best-fit CCSN+PISN model has slightly lower $\chi^2=4.41$ with PISN contributing $\sim 20\hbox{--}30\%$ to elements above Si. However, as can be seen from Fig.~\ref{fig:best-fit_stars}a, the quality of fit is indistinguishable from a single CCSN indicating that including PISN does not improve the quality of fit. 
The best-fit model from 2CCSNe provides a slightly better fit with a $\chi^2=3.95$ with somewhat better fits for O and Zn. However, similar to the single CCSN best-fit model, it gives a poor fit for Ti and all elements from Fe to Cu. 
The quality of fit remains unchanged for the best-fit  CCSN+near-\ch~model as the best-fit corresponds to the combination where there is negligible contribution from SN1a as is evident from $\eta_{\rm 1a}$ from Fig.~\ref{fig:best-fit_stars}a. 
Lastly, for the best-fit CCSN+sub-\ch~model, the quality is marginally better than the best-fit 2CCSNe model with a $\chi^2=3.09$. This model can additionally fit O, Cu, and Zn along with a slightly better fit for Ti at the cost of a poorer fit for V and Mn. 
Surprisingly, the primary reason for the slightly better fit for the best-fit CCSN+sub-\ch~model is that it can perfectly fit Cu even though SN1a contribution to Cu is negligible with $\eta_{\rm 1a}\sim 0.2$ (see the bottom panel in Fig.~\ref{fig:best-fit_stars}a).

We note here that COS171 is not a VMP star. Thus, if the origin of elements in COS171 is entirely from CCSN, it is likely to have come from ejecta from several events. 
However, as mentioned earlier, combining ejecta from multiple sources is not possible as it is computationally expensive. Thus, in order to test whether multiple CCSN ejecta can match the observed abundance pattern we lower the $M{\rm _{dil, CCSN}^{min} }$ to $5\times 10^{3}\,\Msun$ for a single CCSN so that it can match the overall high metallicity in addition to the relative abundance pattern. Interestingly, we find that even with a lower value of minimum dilution, the best-fit model for a single CCSN remains unchanged with a  $M{\rm _{dil, CCSN}=1.1\times 10^4\,\Msun}$ and a $\chi^2=4.59$. The best-fit model for 2CCSNe does slightly better and can additionally fit O and Zn but continues to give a poor fit for Ti and elements from Fe to Cu. The corresponding value of $\chi^2=3.36$ is comparable to the best-fit CCSN+sub-\ch~model.
The reason for the overall poor fit is primarily due to the peculiar abundance pattern from Fe to Zn. In particular, highly sub-solar $\B{Ni}{Fe}\sim -0.7$ along with super-solar $\B{Co}{Ni}\sim 0.5$ cannot be produced by either CCSN or SN1a. 

In summary, even though the best-fit CCSN+sub-\ch~model provides the overall best fit with the lowest $\chi^2$, the quality of fit is comparable to the best-fit 2CCSNe model while the overall quality of fit is poor for both. Thus, the observed abundance pattern in COS171 cannot be strongly attributed to SN1a in sharp contrast to SDSSJ0018-0939. We note, however, that given the relatively high metallicity of COS171 of $\B{Fe}{H}-1.35$ and the fact that it has the highest metallicity among the stars observed in Ursa Minor \citep{cohen2010ApJ}, SN1a could be a natural solution from the point of view of galactic chemical evolution \citep{kirby2019ApJ}. Our analysis, on the other hand, is based entirely on the observed abundance pattern where a clear SN1a signature is absent.

\subsection{BD+80245}
The best-fit for BD+80245 for the different scenarios is shown in Fig.~\ref{fig:best-fit_stars}b with the corresponding details listed in Table~\ref{tab:best-fit_parm_stars}. Similar to COS171, PISN provides an extremely poor fit with $\chi^2=45.7$. The quality of fit for a single CCSN  and 2CCSNe are comparable with best-fit $\chi^2$ of 8.55 and 7.31, respectively, with an overall poor fit as they cannot match the abundance of multiple elements with particularly large deviation from the observed abundances for Mg, Al, Ti, Ni, and Zn. 
The quality of fit from the best-fit CCSN+PISN model is almost identical to the best-fit model from the single CCSN scenario with very little PISN contribution of $\lesssim 5\%$ for Mg and Si and almost near zero contribution from PISN for all other elements. In fact, the CCSN model in the best-fit CCSN+PISN model is also identical to the single CCSN case indicating that no PISN contribution is required to explain the abundance pattern in this star. 
The best-fit CCSN+near-\ch~model provides a similar quality of fit with a $\chi^2=7.63$. The best-fit CCSN+sub-\ch~model gives the lowest $\chi^2=4.18$ even though the overall quality of fit is only slightly better. While it does provide a better fit for Ti, Ni and Zn, the fit is quite poor for Mg, Al, Si, Cr, and Co with large deviations from the observed values well beyond the $1\sigma$ uncertainty.

Because BD+80245 has a metallicity of $\B{Fe}{H}=-1.73$, we also considered the possibility of reducing $M{\rm _{dil,CCSN}^{min}}=5\times 10^3\,\Msun$ similar to COS171. In this case, the quality of fit for the best-fit model from a single CCSN remains almost unchanged with $\chi^2=8.37$, the value is slightly better for the best-fit model from 2CCSNe with $\chi^2=6.58$. Despite the lower $\chi^2$, the quality of fit remains poor as these models cannot match the abundance of any additional elements. The reason for the overall poor fit is primarily due to Al and Mg which are produced exclusively by CCSN. In particular, it has an extremely sub-solar value of $\B{Al}{Fe}\sim -1.7$ leading to highly super-solar $\B{Mg}{Al}\sim 1.2$. This cannot be matched by any of the CCSN models which leads to an overall poor fit for all models. We note that the best-fit PISN model can match the low Al along with the highly super-solar $\B{Mg}{Al}$ but gives an extremely poor fit for almost all other elements.

Thus, similar to COS171, among all the scenarios, although CCSN+sub-\ch~provides the best fit, the quality of fit is poor and it is difficult to claim a clear signature for SN1a contribution for BD+80245 in contrast to SDSSJ0018-0939.

\subsection{HE0533-5340}
The best-fit models for HE0533-5340 are shown in Fig.~\ref{fig:best-fit_stars}c and the details of the models and best-fit parameters are listed in Table~\ref{tab:best-fit_parm_stars}. Similar to other stars, the best-fit PISN model provides an extremely poor fit with a $\chi^2=33.8$. 
Unlike COS171 and BD+80245, the best-fit model from the  CCSN+near-\ch~provides the best overall fit with $\chi^2=3.9$. The primary reason for  CCSN+near-\ch~being the best fit is the elevated $\B{Mn}{Fe}\sim 0.1$ which is a key signature of the near-\ch~model. 
The overall quality of fit, however, is poor as the best-fit model cannot match multiple elements such as Mg, Al, Si, Ti, Cr, and Ni. The best-fit model from CCSN+sub-\ch~ provides a comparable quality of fit with $\chi^2=4.91$. In this case, although the $\chi^2$ is slightly higher for the best-fit CCSN+sub-\ch~model, it can match more elements such as Ti and Ni within the observed $1\sigma$ uncertainty compared to the best-fit  CCSN+near-\ch~model. Interestingly, the best-fit CCSN+PISN model provides a comparable fit with a $\chi^2=4.93$ where PISN contributes to $\sim 50\%$ to the even $Z$ elements from Si to Cr. The quality of fit is slightly better than both the single CCSN and 2CCSNe models that have $\chi^2$ of 5.94 and  5.71, respectively. The best-fit CCSN+PISN and 2CCSNe models can both fit Mg within the $1\sigma$ uncertainty in addition to all the elements that can be fit by the CCSN+near-\ch~model. We note that even though this star can potentially have a PISN origin, it cannot be distinguished from any of the other scenarios.
The overall poor quality of fit for the observed pattern in HE0533-5340 across all scenarios is primarily due to the peculiar abundance features from Mg--Si and Cr--Mn. In particular, similar to BD+80245, the highly sub-solar value of $\B{Al}{Fe}\sim -1.4$ with super-solar $\B{Mg}{Al}\sim 1.1$ cannot be fit by any CCSN model. Additionally, high super-solar $\B{Si}{Fe}\sim 0.5$ cannot be simultaneously fit with extreme sub-solar $\B{Al}{Fe}$. Lastly, sub-solar \B{Cr}{Mn} cannot be produced by either CCSN or SN1a leading to a poor fit for Cr for all models.

In summary, even though  CCSN+near-\ch~provides the best fit, the overall quality of fit remains poor. Additionally, the quality of fit from scenarios that do not include SN1a such as CCSN+PISN and 2CCSNe is comparable and can even match more elements within the observed $1\sigma$ uncertainty. Consequently, the observed abundance pattern in  HE0533-5340 cannot be unambiguously attributed to SN1a.

\begin{figure}
\centering
    \includegraphics[width=\columnwidth]{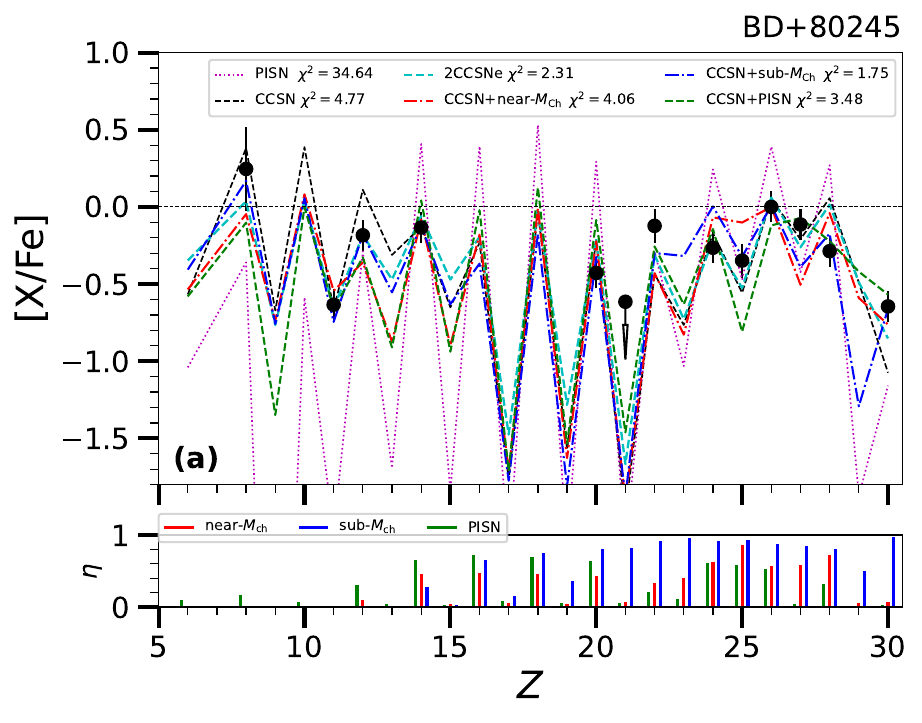}\hfil
    \includegraphics[width=\columnwidth]{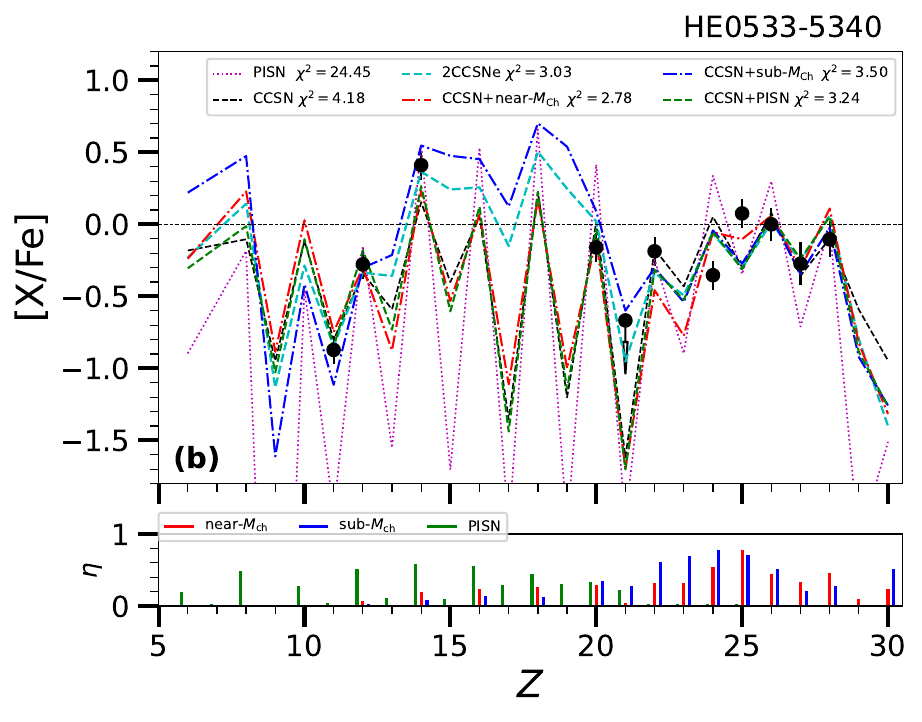}\par\medskip
    \includegraphics[width=\columnwidth]{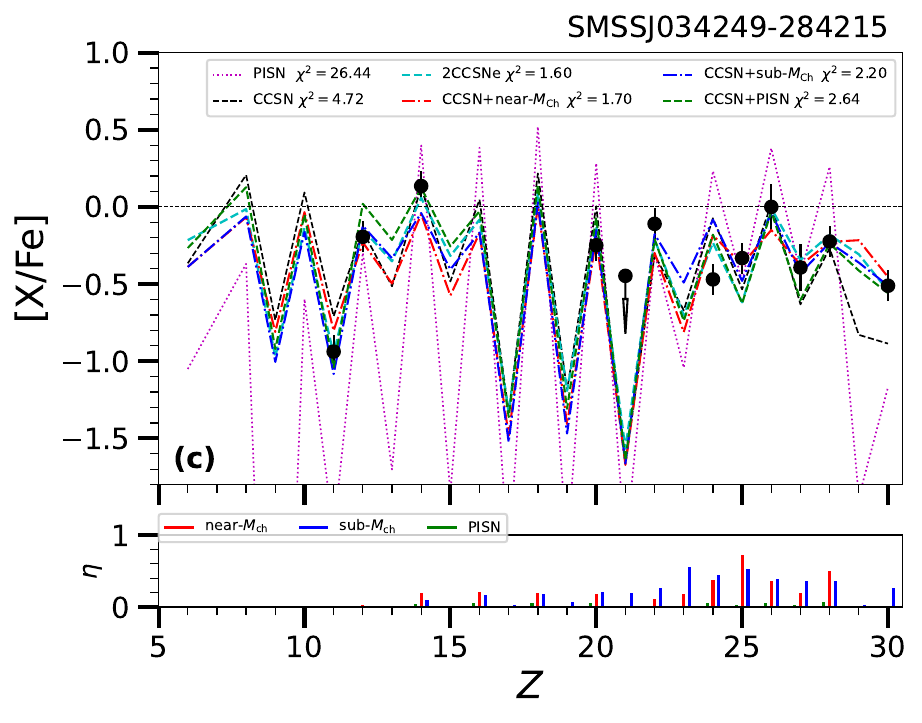}
\caption{Same as Fig.~\ref{fig:best-fit_stars}, but Al is neglected.}
\label{fig:best-fit_noAl}
\end{figure}
\subsection{SMSSJ034249-284216}
The best-fit model for SMSSJ034249-284216 from all the scenarios is shown in Fig.~\ref{fig:best-fit_stars}d. The situation is very similar to COS171, with the best-fit CCSN+sub-\ch~model providing the overall best fit with $\chi^2=4.13$ and a comparable quality of fit from the best-fit 2CCSNe model with $\chi^2=4.73$. Among these two models, the best-fit CCSN+sub-\ch~model can match the abundance of Ca, Ti, and Mn whereas the best-fit 2CCSNe model can match Si and Zn and vice-versa. Compared to the best-fit CCSN+sub-\ch~model, the best-fit models from a single CCSN and  CCSN+near-\ch~provide a slightly worse fit whereas a single PISN provides an extremely poor fit. Similar to HE0533-5340, the best-fit CCSN+PISN model can provide a fit comparable to the 2CCSNe case and has major contributions of up to $\sim 80\%$ to most elements from Mg to Ni with a CCSN model that is distinct from both the single CCSN and 2CCSNe models. This star can potentially have a PISN origin, which, however, cannot be distinguished from other scenarios as is the case for HE0533-5340 discussed above.
The overall quality of fit from all scenarios is poor which is primarily because they completely fail to match the abundances of Mg, Al and Cr. Similar to BD+80245, this star has an extremely sub-solar value of $\B{Al}{Fe}\sim -1.4$ leading to highly super-solar $\B{Mg}{Al}\sim 1.2$ that cannot be matched by any of the CCSN models. Similar to HE0533-5340, sub-solar \B{Cr}{Mn} cannot be produced by any source yielding a poor fit for Cr for all models.   

Thus, the situation is similar to COS171 where although the best-fit CCSN+sub-\ch~model provides the overall best fit with the lowest $\chi^2$, the quality of fit is comparable to the best-fit 2CCSNe and CCSN+PISN models while the overall quality of fit is poor for both. Thus, the observed abundance pattern in SMSSJ034249-284216 cannot be strongly attributed to SN1a.

\subsection{Best-fit without Al for BD+80245, HE0533-5340, and SMSSJ034249-284216}
As discussed in the above analysis, the extremely sub-solar value of $\B{Al}{Fe}$ is common to all three stars and is one of the primary reasons for the overall poor fit. We note, however, that a substantially higher abundance of $\B{Al}{Fe}$ has been reported for both BD+80245~\citep{ivans2003ApJ} and HE0533-5340~\citep{cohen_2013} with $\Delta \B{Al}{Fe}$ of $+0.41$ and $+0.68$ compared to \citet{reggiani2023AJ}. Additionally, \citet{Marino2019MNRAS} has found similar values of $\B{Al}{Fe}$, they have reported extremely high $\sigma$ of $0.49$ for SMSSJ034249-284216. 
Because the Al abundance is potentially uncertain and given that it has a large effect on the quality of fit, we repeat the analysis for the three stars by neglecting Al. 
The best-fit results are presented in Fig.~\ref{fig:best-fit_noAl}. Overall, neglecting Al does lead to a substantial improvement in the quality for all stars.  As before, the overall best-fit model for BD+80245 and HE0533-5340 comes from CCSN+sub-\ch~ and  CCSN+near-\ch, respectively, with much improved $\chi^2$ of 1.75 and 2.78, respectively. However, in both cases, the quality of fit from the best-fit 2CCSNe model is comparable with $\chi^2$ of 2.31 and 3.03 for BD+80245 and HE0533-5340, respectively. Interestingly, for SMSSJ034249-284216, the overall best-fit model is from 2CCSNe with a $\chi^2$ of 1.6 which is closely followed by  CCSN+near-\ch with $\chi^2$ of 1.7. 

Thus, in summary, the results clearly show that the abundance pattern in  BD+80245, HE0533-5340, and SMSSJ034249-284216 cannot be strongly associated with SN1a as ejecta from CCSNe alone can also provide comparable fit.

\section{Summary and Conclusions}\label{sec:summary}
SDSSJ0018-0939 is an $\alpha$-poor VMP star that was previously identified as a star with a possible signature from PISN by \citet{aoki2014Sci}. However, we find that none of the theoretical PISNe abundance patterns is remotely compatible with the observed pattern as the best-fit PISN pattern gives an extremely poor fit with a high $\chi^2$ which is consistent with the results found by~\citet{takahashi2018ApJ}. Even when CCSN+PISN scenario is considered, the best-fit solution is found to be the one with zero contribution from PISN which indicates that the abundance pattern does not have any PISN features.
Instead, the peculiar abundance pattern observed in SDSSJ0018-0939 can be fit perfectly by the mixing of the ejecta between a $30\,\Msun$ CCSN with negligible fallback and a sub-\ch~SN1a resulting from He detonation of a WD with $M_{\rm CO}=1\,\Msun$ and $M_{\rm He}=0.05\,\Msun$. In particular, we find that the unique pattern around the Ti-Cr region with $\B{X}{Fe}\gtrsim 0$ for Ti, V, and Cr along with $\B{Ti}{Cr}\gtrsim 0$ is the key signature of  sub-\ch~SN1a resulting from contribution from He shell burning that is distinct from any other source such as CCSNe or a combination of CCSN and near-\ch~SN1a. This, in combination with very low $\B{C}{Fe}\sim -1$ along with highly sub-solar values of $\B{X}{Fe}$ for Mg, Si, and Ca, point to a near smoking-gun signature of a star formed from a gas strongly polluted by a sub-\ch~SN1a. 

A clear signature of a sub-\ch~SN1a in the early Galaxy is particularly interesting as it is consistent with the findings from recent studies that analyzed the observed evolution of $\B{Mn}{Fe}$ and $\B{Ni}{Fe}$ in the Milky Way and dwarf galaxies \citep{seitenzal2013A&A,kobayashi2020ApJ,eitner2020,eitner2023}. These studies conclude that the contribution of sub-\ch~SN1a is crucial and could account for up to $\sim 50\hbox{--}75\%$ of the total SN1a. Furthermore, there are indications that sub-\ch~SN1a is dominant in the early times in dwarf galaxies whereas near-\ch~SN1a becomes important at later times \citep{kirby2019ApJ,reyes2020ApJ}. Thus, it is not surprising that a signature from a sub-\ch~SN1a is more likely to be found in VMP stars.  

We also analyzed four stars, namely, COS171, BD+80245, HE0533-5340, and SMSSJ034249-284216, that have been previously identified as stars that show a possible signature of SN1a \citep{mcwilliam2018ApJ,reggiani2023AJ}. Overall, we find that none of the scenarios can provide a good fit to the observed abundance pattern in these four stars. We find that even though the best-fit models from scenarios involving CCSN+SN1a provide the best fit with the lowest $\chi^2$, the best-fit models from 2CCSNe can always provide comparable fits with a marginally higher $\chi^2$. Interestingly, we also find comparable fits with CCSN+PISN scenario with a substantial contribution from PISN for HE0533-5340 and SMSSJ034249-284216 although it is indistinguishable from other best-fit scenarios. Thus, although the elements observed in these stars could have originated from SN1a, the fact that CCSN ejecta or a combination of CCSN and PISN ejecta can match the abundance pattern almost equally well implies that the origin of elements in these stars cannot be unambiguously attributed to SN1a. The conclusion remains unchanged even when Al is neglected for the analysis even though the quality of fit improves substantially for BD+80245, HE0533-5340, and SMSSJ034249-284216. In this case, the quality of fit from the best-fit 2CCSNe model is again comparable to the best-fit CCSN+SN1a models for  BD+80245 and HE0533-5340. On the other hand, the best-fit 2CCSNe model provides the overall best fit for SMSSJ034249-284216. Thus, none of the four stars has a clearly distinguishable SN1a signature in sharp contrast to SDSSJ0018-0939 where a clear signature of an SN1a is established from the fact that a near-perfect fit is achieved only from the best-fit CCSN+sub-\ch~SN1a whereas all other scenarios fail to provide a good fit. 

The significance of SDSSJ0018-0939 as the star with the clearest signature of SN1a among all currently known VMP stars is ultimately based on the robustness of observed abundances of key elements such as C, Ti, V, and Cr. In this regard, we note that large discrepancies in the estimated abundances of elements are sometimes found when a given star is analyzed by different groups. Large changes in the abundance of key elements can consequently affect and even change their association with a particular source. A recent example of this is the chemically peculiar star LAMOST J1010+2358 that was first reported by \citet{xing2023Natur} whose estimated abundance pattern is found to be considerably different in subsequent analysis by \citet{skuladottir2024ApJ} and \citet{Thibodeaux2024}. Thus, it is important that SDSSJ0018-0939 is also analyzed by independent groups to confirm whether the observed abundance pattern is robust. 

Our results highlight the importance of $\alpha$-poor VMP stars as powerful probes for gaining crucial insights about SN1a. Such stars provide the only direct probe for detailed nucleosynthesis in SN1a that can differentiate between near-\ch~and sub-\ch~models. However, we also find that not all $\alpha$-poor VMP stars require SN1a contribution as some of the CCSN models that do not undergo substantial fallback can also match the observed abundance pattern equally well. Thus, a clear association of SN1a with such stars should only be made after careful analysis. Notwithstanding this, future detections of more $\alpha$-poor VMP stars with detailed abundance patterns from multiple elements are required to find more stars such as SDSSJ0018-0939 that can be used to gain crucial insights into SN1a and their contribution to the early Galaxy.

\section*{Data Availability}
Data is available upon reasonable request.

\bibliography{main}{}
\bibliographystyle{aasjournal}



\end{document}